# Long-term spin state storage using ancilla charge memories

Harishankar Jayakumar[1], Artur Lozovoi[1], Damon Daw[1], and Carlos A. Meriles[1,2,*]

[1]*Department. of Physics, CUNY-City College of New York, New York, NY 10031, USA.*
[2]*CUNY-Graduate Center, New York, NY 10016, USA.*
[*]*Corresponding author. E-mail: cmeriles@ccny.cuny.edu*

We articulate confocal microscopy and electron spin resonance to implement spin-to-charge conversion in a small ensemble of nitrogen-vacancy (NV) centers in bulk diamond, and demonstrate charge conversion of neighboring defects conditional on the NV spin state. We build on this observation to show time-resolved NV spin manipulation and ancilla-charge-aided NV spin state detection via integrated measurements. Our results hint at intriguing opportunities in the search for enhanced forms of color-center-based metrology and information processing down to the limit of individual point defects.

Paramagnetic color centers in semiconductors are presently capturing widespread interest as versatile qubits for nanoscale sensing and quantum information processing[1,2]. Optical access to individual color centers coupled to single electron spin control and millisecond-long coherence spin lifetimes have led to stunning demonstrations of entanglement and quantum logic in diamond and other wide-bandgap materials[3,4]. Although the prevalent approach in much of this work is almost exclusively focused on the relationship between the defect spin state and the photons it emits or absorbs, recent efforts have been expanded to explore the virtual atom's 'valence' charge carriers as a means for transporting quantum information[5], creating local environments protected against spin decoherence[6], or enabling more convenient forms of defect spin readout[7]. For example, the use of spin-to-charge conversion schemes has led to demonstrations of electrical spin qubit detection in diamond[8-10]. Another intriguing possibility is the use of spin-selective defect ionization as a route for spin-polarized carrier injection, of interest for spintronics applications in systems where the use of ferromagnetic electrodes is impractical[5].

Here we combine magnetic resonance and multi-color confocal microscopy to alter the charge state of a small ensemble of negatively charged nitrogen-vacancy (NV⁻) centers in diamond conditional on their spin state. Successive cycles of ionization and recombination of the 'qubit' NVs respectively produce free electrons and holes, which we subsequently capture via an ensemble of neighboring carrier-type-selective traps. Adapting this ancilla-aided integrated detection (AID) strategy to time resolved measurements, we demonstrate basic building blocks of NV⁻ spin control for different trap types. Although under the present conditions standard photo-luminescence readout still proves more sensitive, we expose through experiment and modeling a broad parameter space, that could be potentially exploited not only to boost sensitivity beyond existing techniques, but also, more generally, as a platform for applications where the charge carrier itself serves as a flying qubit.

The cartoon in Fig. 1a summarizes our starting working geometry, comprising small ensembles of NV centers surrounded by a larger set of charge traps[11]. The latter take the form of point defects whose fluorescence changes (e.g., from dark to bright or vice versa) upon capture of a carrier. For the present studies, we first use silicon-vacancy (SiV) centers, whose charge state can be controlled with light pulses of suitable wavelength[12,13]. Though surface effects[14-17] and/or carrier tunneling[18] can render the charge state of defects unstable, SiV centers at moderate concentrations (~0.1 ppm in the present sample) are known to feature long-lived charge states and hence serve as classical memories with virtually unlimited storage time in the dark[19].

The mechanism underlying spin-conditional ionization of the NV — a spin $S = 1$ defect featuring triplet ground and excited states[20] — can be understood with the help of the energy diagram in Fig. 1b. Optical excitation induces a spin-preserving transition within the triplet manifold, followed by radiative decay. Inter-system crossing to a manifold of intermediate singlet states (where $S = 0$) is more likely when the initial NV triplet state is $|m_S = \pm 1\rangle$, thus leading to spin-selective shelving (the basis for NV spin optical readout[21]). At sufficiently high laser powers, photon absorption during the excited triplet lifetime (~10 ns) propels the NV⁻ excess electron into the conduction band, hence changing the charge state of the NV into neutral[22]. When the illumination interval is comparable with the shelving time (~100 ns), the above two-step photon process is more probable for the $|m_S = 0\rangle$ state — comparatively immune to intersystem crossing — thus leading to spin-selective electron injection into the conduction band. This spin-to-charge conversion (SCC) process[7,23-25] underlies recent demonstrations of electrical spin readout[8,9], down to individual NVs[10].

In our experiments, we first study a [100] type 1b diamond crystal simultaneously hosting NVs, SiVs, and nitrogen impurities with approximate concentrations of $10^{-2}$ ppm, $10^{-1}$ ppm, and 1 ppm, respectively. Figures 1c through 1e lay out the fundamentals of our spin storage protocol: We use multiple red (632 nm) laser scans to charge-initialize NVs and SiVs within a $40 \times 40$ μm² area into a non-fluorescent state[26]. We then cycle the NVs at the center point in this region between their neutral and negative charge states via simultaneous green (520 nm) and red laser pulses (100 ns). The latter are separated by (optional) microwave (MW) pulses, whose duration (100 ns) and frequency (2.87 GHz) are adjusted so as to invert the populations of the $|m_S = 0\rangle$ and $|m_S = \pm 1\rangle$ states of the ground triplet (Fig. 1c). Each event of NV ionization (i.e., NV⁻→NV⁰) and recombination (i.e., NV⁰→NV⁻) respectively results in the generation of a free electron and a hole, which subsequently diffuse away from the illumination point to be ultimately captured by a neighboring trap. Under these conditions, SiVs exhibit a one-way transformation into the negatively charged, bright state[26], thus leading to the formation of a fluorescent disk centered around the point of optical excitation (Fig. 1d). Note that identical numbers of electrons and holes are injected during multiple repetitions of the qubit control protocol, hence allowing one to store the qubit spin state via the capture of one carrier type or the other, provided the traps are predominantly sensitive to one type of carrier, the case for SiV[26].

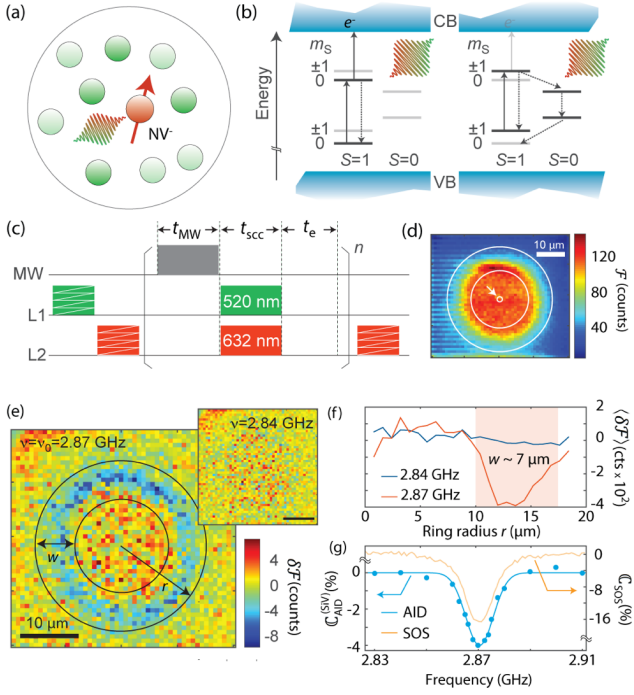

**FIG. 1. Charge storage of NV⁻ spin state.** (a) Ancilla charge traps change their fluorescence upon capture of carriers produced by NV ionization/recombination. (b) Schematics of NV charge photoionization dynamics. (c) Experimental protocol. The green (red) laser power is 3 mW (21 mW) and zig-zags denote charge initialization and readout raster scans; $t_e = 1$ μs is the wait time between cycles. (d) SiV⁻–selective fluorescence image $\mathcal{F}$ upon use of the protocol in (c) for a total time $t = 10$ s. The white arrow points toward the illuminated area; the medium and outer circles demark the region of maximal spin contrast. (e) Differential SiV⁻ fluorescence patterns $\delta\mathcal{F} \equiv \mathcal{F}_{on} - \mathcal{F}_{off}$; subscripts refer to the presence or absence of MW resonant with (2.87 MHz) or far detuned from (2.84 MHz) the NV⁻ ground-state crystal-field frequency. (f) SiV⁻ differential fluorescence averaged over a 1-μm-wide ring of variable radius $r$ for the on- and off-resonance images in (e). (g) NV⁻ magnetic resonance spectrum using SiV-AID or SOS.

For $n \gg 1$ cycles of ionization–recombination, the integrated number of carriers — and, correspondingly, the average radius of the SiV⁻ disk in the ensuing confocal image— is larger when the MW field is off (because the NV⁻ spin state is $|m_S = 0\rangle$, where ionization is more likely). We expose this spin-dependent contrast in Fig. 1e, where we subtract the SiV fluorescence patterns obtained with and without resonant MW acting on the NV crystal-field transition. This trap-encoded spin signal (SiV-AID) takes the form of a concentric dark ring, absent when the MW is detuned off resonance (left and right panels in Fig. 1e, respectively).

To quantify the effect at arbitrary MW frequencies $\nu$, we first calculate the integrated SiV⁻ fluorescence $\langle\mathcal{F}\rangle(r_i, \nu) = \sum_j \mathcal{F}(\varphi_j, r_i, \nu)$ over all angles $\varphi_j$ along concentric, 1-μm-wide rings of increasing radius $r_i$. We subsequently use this radial distribution to determine the contrast $\mathbb{C}_{AID}^{(SiV)}(\nu) = 2(I_{on}(\nu) - I_{off})/(I_{on}(\nu_0) + I_{off})$, where $I(\nu) = \sum_i \langle\mathcal{F}\rangle(r_i, \nu)$ is the integrated fluorescence, here restricted to a ring of width $w \approx 7$ μm around $r \approx 14$ μm for optimal SNR (see Fig. 1f). In the above expression, $\nu_0$ is the NV⁻ spin resonance frequency, and the label indicates on/off MW. Using these definitions, Fig. 1g shows the SiV-AID magnetic resonance spectrum of NV⁻. We find it to be in good agreement with that obtained from collecting the NV⁻ spin-dependent photoluminescence, the most common readout protocol here referred to as standard optical sensing or SOS.

Given the unlimited lifetime of trapped charge states at these defect concentrations[19, 27], signal integration can be carried out over a broad time interval, here ranging from 50 ms to 2 s. Figs. 2a and 2b respectively show some snapshots of the evolving SiV charge distribution within this temporal span, alongside the integrated spin signal at each of these times. Further, the present approach can be easily adapted to time resolved measurements of the spin qubit. A first demonstration is presented in Fig. 2c where we use the sequence in Fig. 1c with a variable MW pulse duration to measure the NV⁻ spin Rabi response. A more involved scheme is shown in Fig. 2d, this time adapted to measure the NV⁻ spin echo signal (Fig. 1e).

While SiVs serve as convenient traps for the present application, they are certainly not the only type of defects one can resort to. One immediate possibility are NV centers, effectively trapping holes when negatively charged, but displaying poor electron capture cross section in the neutral state[26]. Extending the results above, here we implement NV-encoded spin storage using ancilla NVs sufficiently removed from the point of laser illumination. Fig. 3a shows the experimental protocol: Unlike in Fig. 1c, in this case we use a green laser scan to charge-initialize the area around the point of optical excitation into a majority of negatively charged NVs. For these experiments, we use a SiV-free diamond crystal with nitrogen and NV concentrations of 1 ppm and $10^{-2}$ ppm, respectively. Though convenient to enhance sensitivity, this modified diamond composition is not mandatory, as the SiV⁻ fluorescence can be selectively filtered out from the recorded emission spectrum[26]. Fig. 3b displays the fluorescence $\mathcal{G}$ from ancilla NVs emerging after multiple repetitions $n$ of the SCC protocol and a red readout scan: Surrounding the central bright spot, we observe the formation of a pronounced dark halo, indicative of ancilla transformation to a majority neutral charge state via the capture of holes diffusing from the point of illumination (inner white circle in Fig. 3b).

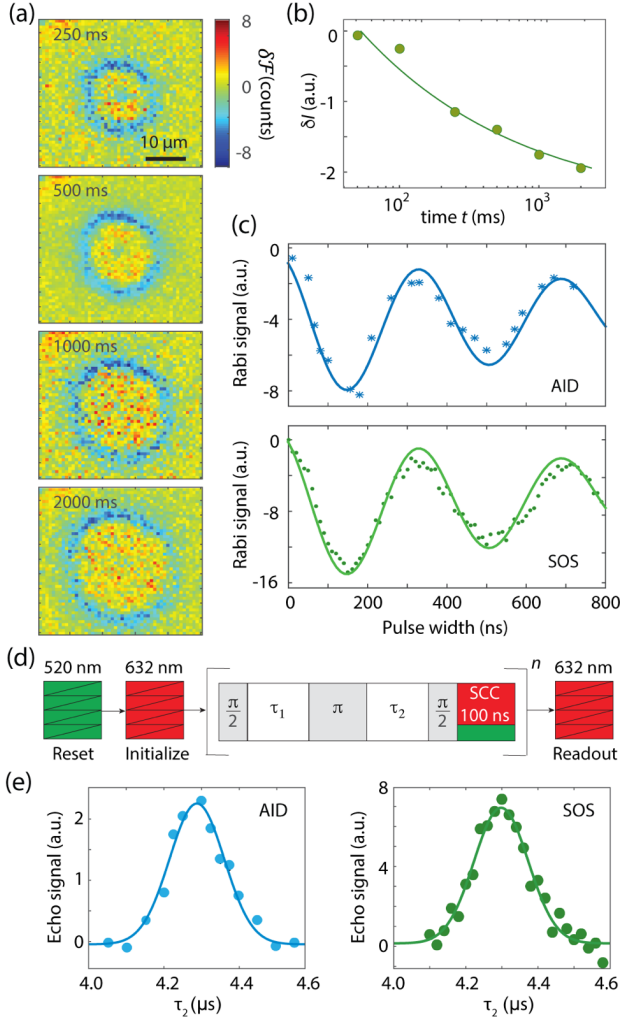

**FIG. 2. Time-resolved, charge-encoded NV⁻ spin dynamics.** (a) SiV⁻ differential charge patterns for different integration times. (b) On-resonance SiV-AID signal $\delta I(\nu_0)$ as a function of time. (c) NV⁻ spin Rabi signal using the protocol in Fig. 1c for resonant MW pulses of variable duration or using SOS readout (upper and lower plots, respectively). (d) AID-adapted NV⁻ spin-echo sequence; SCC denotes spin-to-charge conversion using simultaneous, 100-ns-long red and green optical pulses, as in Fig. 1c. (e) Measured AID (left) and SOS (right) Hahn-echo signals of the NV spin. In all plots, solid traces are guides to eye.

To reveal the spin state of the qubit NVs (i.e., the NVs directly exposed to the green/red beams) we subtract the fluorescence patterns $\mathcal{G}_{on/off}$ — this time produced by the ancilla NVs — with and without MW. We find a non-zero difference only when the microwave is resonant with the NV⁻ crystal field transition (left image in Fig. 3c), thus yielding an NV-encoded integrated spin signal (NV-AID). Contrary to Fig. 1e, the fluorescence change $\delta\mathcal{G} \equiv \mathcal{G}_{on} - \mathcal{G}_{off}$ is positive, corresponding to less ancilla NVs transitioning to neutral when the MW is on; the latter, in turn, agrees with the notion of less frequent NV⁻ ionization/recombination cycles due to MW-induced NV⁻-state shelving in the singlet manifold. Interestingly, the maximum fluorescence difference is found near the point of illumination, indicative of a spin-dependent NV⁻ concentration under optical excitation. Note, however, this contribution has a negligible impact on the radial fluorescence distribution $\langle\delta\mathcal{G}\rangle$ — vanishing near the origin, Fig. 3d — and thus on the integrated NV-AID signal $\delta J(\nu) = \sum_i \langle\delta\mathcal{G}\rangle(r_i, \nu)$. As a matter of fact, we measure a time growth (Fig. 3e) comparable to that observed above for silicon-vacancies (Fig. 2b). Likewise, we attain good agreement between the NV-AID and SOS spectra (Fig. 3f).

To expose the range of conditions where the use of AID can be advantageous, we model spin-to-charge conversion via the stochastic variables $q_j$, $j = 0, 1$, respectively associated with the probabilities of generating a charge carrier when the initial NV⁻ state has spin projection $|m_S| = j$. Using $\langle\rangle$ and $\delta^2$ to respectively denote mean values and variances, we find that the detection sensitivity after $n$ repeats is given by[11]

$$\eta_{AID} = \frac{\sqrt{t_{AID}}\sqrt{\sum_j [\delta^2(k_a \nu_j) + (1-\langle\lambda\rangle)\langle k_a\rangle^2 \langle \nu_j\rangle^2]}}{\sqrt{\langle\lambda\rangle}\langle p\rangle|\langle q_0\rangle - \langle q_1\rangle|\langle k_a\rangle}. \quad (1)$$

where $\langle\lambda\rangle$ is the fraction of carriers captured by the ensemble of ancilla traps, $\langle k_a\rangle$ is the average number of photons collected during charge readout of an ancilla, and $\nu_j \equiv pq_j + w$ is a stochastic variable describing ancilla trap activation. In the latter expression, $w$ denotes contributions from background carriers (i.e., produced by the ionization of defects other than the qubit NV[11]), and $p$ is a Boolean stochastic variable associated with the probability of finding the NV in the negatively charged state prior to SCC. Finally, $t_{AID} = t_{scc} + t_e + (t'_{ia} + t'_{ra})/n$ is the average time per repeat during AID, calculated as the sum of contributions from the SCC light pulse duration $t_{scc}$, the spin evolution time $t_e$ during the chosen protocol, and the charge initialization and readout times of the ancilla trap ensemble, $t'_{ia}$ and $t'_{ra}$, respectively.

Depending on the fidelity of the SCC process and the ancilla brightness, we identify different, complementary regimes: For example, assuming, for simplicity, $\langle\lambda\rangle \sim \langle p\rangle \sim 1$ and in the limit where $\langle k_a\rangle^2 \sum_j \delta^2 \nu_j \gg \langle k_a\rangle \sum_j \langle\nu_j\rangle^2$ (typically corresponding to $\langle k_a\rangle \gg 1$), we obtain

$$\eta_{AID} \sim \frac{\sqrt{t_{AID}}\sqrt{\langle q_0\rangle(1-\langle q_0\rangle) + \langle q_1\rangle(1-\langle q_1\rangle)}}{|\langle q_0\rangle - \langle q_1\rangle|}, \quad (2)$$

dependent only on the qubit spin-to-charge conversion probabilities. Conversely, when $\langle k_a\rangle^2 \sum_j \delta^2 \nu_j \ll \langle k_a\rangle \sum_j \langle\nu_j\rangle^2$ the sensitivity can be written as

$$\eta_{AID} \sim \frac{\sqrt{t_{AID}}\sqrt{\langle q_0\rangle^2 + \langle q_1\rangle^2}}{|\langle q_0\rangle - \langle q_1\rangle|\sqrt{\langle k_a\rangle}}, \quad (3)$$

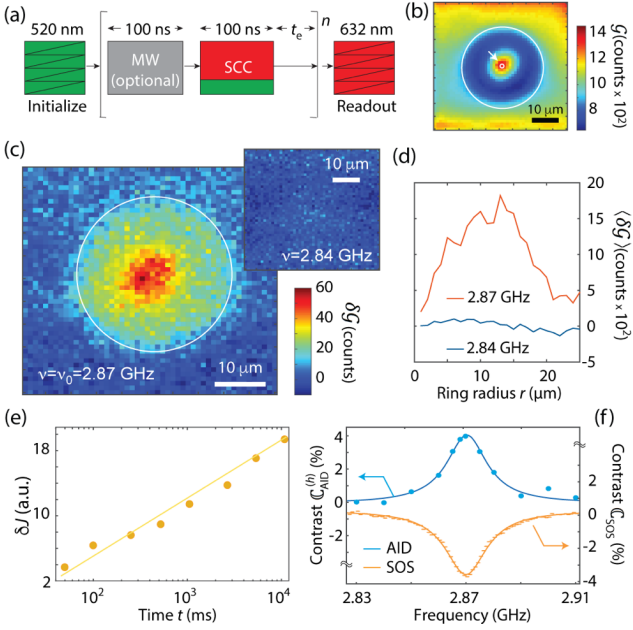

**FIG. 3. NV-aided long-term spin state storage.** (a) Experimental protocol; $t_e = 1$ μs is the wait time between successive cycles. (b) NV$^-$ fluorescence image using the scheme in (a). Excluding the initialization and readout scans, the total signal integration time amounts to 10 s. (c) NV$^-$ differential fluorescence images $\delta\mathcal{G} \equiv \mathcal{G}_{on} - \mathcal{G}_{off}$ using resonant (main) and off-resonant (insert) MW. (d) NV$^-$ differential fluorescence averaged over 1-μm-wide rings of variable radii $r$ for resonant or detuned MW (red and blue traces, respectively). (e) On-resonance NV-AID signal $\delta J(\nu_0)$ as a function of time. (f) NV$^-$ spin resonance spectra as determined using NV-AID or SOS (blue and yellow dots, respectively). In all plots, solid traces are a guide to the eye.

improving inversely with the SCC contrast and the square root of the number of photons collected during ancilla charge readout. Both limits are formally identical to those found in standard SCC-based detection, implying that AID shares the benefits of charge readout[24,25]. Importantly, however, the integrated nature of the AID detection removes the characteristic SCC time overhead so long as the ensemble initialization and readout times are sufficiently short. Further, in the regime of Eq. (3), additional gain over SCC can be attained by choosing ancilla traps featuring photon emission rates exceeding that of the qubit, and/or with longer ionization/recombination times[11]. For an NV center acting as the spin qubit, this latter limit can be attained at low temperatures where optical transitions are spin selective[28] (thus facilitating efficient spin-conditional charge initialization and ionization, i.e., $\delta^2\nu_j\sim0$).

While the present conditions are unfavorable, there is much room for improvement as the location, concentration, and type of ancilla defects — here simply defined by the intrinsic conditions of our samples during crystal growth — can instead be separately optimized using existing defect engineering protocols. Future work can benefit from defect engineering and charge guiding, e.g., to suppress background carriers produced either during NV charge initialization, or from ionization of other coexisting defects near the qubit site. By the same token, arrays of electrodes could be exploited to prevent unintended carrier trapping away from the target ancillae, or to physically separate electrons and holes so as to double the integrated signal. Along the same lines, we note that because photo-generated carriers diffuse in three dimensions (3D), the two-dimensional (2D) plots observed here contain only a fraction of the total spin signal.

Interestingly, AID and SOS are not mutually exclusive as the latter can easily be made part of the former by recording the NV fluorescence during (a fraction of) the SCC pulse. By the same token, since the nuclear spin state of the host nitrogen is robust to NV$^-$ ionization[19], repetitive nuclear spin readout schemes[29,30] can be adapted so as to enhance the number of carriers produced during each repeat via multiple runs of the SCC pulse in each repeat.

Given the ubiquity of defect ionization/recombination and carrier trapping, the ideas herein can be generalized to other qubit–ancilla pairs in diamond or other wide-bandgap semiconductors. Reading the spin state of the qubit defect through the fluorescence of an ancilla emitter separates spin qubit manipulation and readout into two independent processes that can be individually optimized. The latter provides an intriguing route to more efficiently probe spin qubits with photon emission at inconvenient wavelengths (such as the SiV$^0$ in diamond), with a low quantum yield (such as rare earth ions), or with undesired phonon-shifted fluorescence (such as the NV).

Besides sensing, our observations could be further explored more broadly as an alternative route towards the use of charge carriers to transport quantum information. For example, spin-to-charge conversion fidelities reaching nearly 100% can be arguably reached at low temperatures, where the NV photon absorption is spin-dependent[31]. Therefore, resonant ionization of a spin qubit previously entangled with the nuclear spin of its host can lead to a state where the presence or absence of a carrier quantum-correlates with the spin state of the nuclear register in the source qubit. In this context, the question as to whether such carriers can be used as a quantum bus to communicate remote spin qubits is an intriguing one, recently considered theoretically for spin-polarized electrons injected upon ionization of P1 centers[5]. Interestingly, preserving the carrier spin state during transport — as required in Ref. [5] — would not be mandatory in the present case, even though the possibility of injecting spin-polarized carriers in diamond is attractive in its own right.

**Acknowledgments**. We thank J. Henshaw for fruitful discussions. The authors acknowledge support from the National Science Foundation through grants NSF-1619896, NSF-1726573, NSF-1914945, and from Research Corporation for Science Advancement through a FRED Award; they also acknowledge access to the facilities and research infrastructure of the NSF CREST IDEALS, grant number NSF-HRD-1547830.

Supplementary Material for

# Long-term spin state storage using ancilla charge memories

**Harishankar Jayakumar[1], Artur Lozovoi[1], Damon Daw[1], and Carlos A. Meriles[1,2,*]**

[1]Department. of Physics, CUNY-City College of New York, New York, NY 10031, USA.
[2]CUNY-Graduate Center, New York, NY 10016, USA.

**Supplementary Note 1 | Experimental**
The sample used to collect data presented in Fig.1-2 is a <100> CVD grown diamond with defect concentrations of $10^{-2}$ ppm, $10^{-1}$ ppm and 1 ppm NVs, SiVs and substitutional nitrogen impurities, respectively. The sample used to collect data presented in Fig. 3 is a <100> CVD grown diamond with defect concentrations of $10^{-2}$ ppm NVs and 1 ppm substitutional nitrogen (no detectable SiVs). Both samples were purchased from Delaware Diamond Knives. We used a home-built confocal microscope with an air objective (NA=0.7), 520 nm and 632 nm diode lasers with ~1-µm-diameter illumination spots, and a single photon counting module for PL detection. The red laser power can be adjusted on the fly for initialization, SCC, or readout. For NV (SiV) fluorescence readout, we use a passband filter transparent in the spectral window 650 nm to 800 nm (710 nm to 770 nm). We apply microwaves through the use of an omega shaped stripline (0.5 mm diameter) imprinted on a board, in turn serving as the diamond supporting substrate. Note that the beam diameter is much smaller than the carrier diffusion area (reaching a diameter of up to ~25 µm). Therefore, with the exception of a negligible fraction at or near the beam spot, ancilla NVs remain largely unpolarized and hence are on average insensitive to the action of microwave pulses.

**Supplementary Note 2 | Calculation of detection sensitivity**

In this section, we derive Eqs. (1) through (3) in the main text; we also explore specific regimes of these expressions not explicitly considered in the main narrative. For clarity, we start by considering standard optical sensing (SOS) of the qubit, and we subsequently adapt the formalism to spin-to-charge conversion and ancilla-aided detection (AID).

*i-SOS sensitivity*
This sensing modality uses fluorescence readout of the spin qubit during a time $t_r$ shorter than that required for qubit initialization, $t_i$. We start by modeling the qubit as a two-level system with states $|0\rangle$ and $|1\rangle$. The SOS measurement operator is

$$M_{\text{SOS}} = k_0 |0\rangle\langle 0| + k_1 |1\rangle\langle 1|, \qquad (S1)$$

where $k_0$ and $k_1$ are stochastic variables indicating the number of detected photons in each spin state. We model each variable via probability distributions

$$\mathcal{P}_{k_0}(m,t) = \frac{(K_0(t))^m}{m!} e^{-K_0(t)}, \qquad (S2)$$

and

$$\mathcal{P}_{k_1}(m,t) = \frac{(K_1(t))^m}{m!} e^{-K_1(t)}, \qquad (S3)$$

---
[*] E-mail: cmeriles@ccny.cuny.edu

where $K_0(t) = \int_0^t \kappa_0(t')dt'$ and $K_1(t) = \int_0^t \kappa_1(t')dt'$ with $\kappa_0(t')$, $\kappa_1(t')$ denoting spin-dependent photon emission rates satisfying $\kappa_0(t') \sim \kappa_1(t')$ when $t' \geq t_r$. For the case of a single NV⁻ qubit, $t_r \sim 300$ ns for regular laser illumination intensities.

In a typical SOS experiment, the qubit undergoes multiple cycles $n$ of a protocol comprising time intervals of spin initialization $t_i$, evolution $t_e$ (dependent on the application at hand), and readout $t_r$ hence demanding a time per repeat equal to $t_{SOS} = (t_i + t_e + t_r)$,. The observed SOS signal can then be cast as

$$S_{SOS}(t_e, t_r) = \langle Tr\{M_{SOS}(t_r)(U(t_e)\rho(0)U^\dagger(t_e) - \rho(0))\}\rangle, \tag{S4}$$

where the outer brackets indicate ensemble average, and $U(t_e)$ is the time evolution operator. Assuming $\rho(0) = |0\rangle\langle 0|$ and writing $\rho(t_e) \equiv U(t_e)\rho(0)U^\dagger(t_e) = |\psi\rangle\langle\psi|$ with $|\psi\rangle = u_0|0\rangle + u_1|1\rangle$, we find

$$S_{SOS}(t_e, t_r) = |u_1|^2 |\langle k_0\rangle(t_r) - \langle k_1\rangle(t_r)|, \tag{S5}$$

where $\langle k_0\rangle(t_r) = \sum_m m \, \mathcal{P}_{k_0}(m, t_r)$ and $\langle k_1\rangle(t_r) = \sum_m m \, \mathcal{P}_{k_1}(m, t_r)$ represent the average number of photons emitted by the qubit in the $|0\rangle$ and $|1\rangle$ states, respectively.

Since observations are 'differential' (i.e., rely on a reference measurement without spin evolution), we write the signal variance as

$$\delta^2 S_{SOS}(t_e, t_r) = \langle Tr\{M_{SOS}^2(t_r)(\rho(t_e) + \rho(0))\}\rangle - \langle Tr\{M_{SOS}^2(t_r)\rho(t_e)\}\rangle^2 - \langle Tr\{M_{SOS}^2(t_r)\rho(0)\}\rangle^2. \tag{S6}$$

After some algebra, we find

$$\delta^2 S_{SOS}(t_e, t_r) = |u_0|^2 \delta^2 k_0 + |u_1|^2 \delta^2 k_1 + \delta^2 k_0 \geq \delta^2 k_0 + \delta^2 k_1, \tag{S7}$$

where we defined $\delta^2 k_j \equiv \langle k_j^2\rangle - \langle k_j\rangle^2$, $j = \{0,1\}$. Therefore, the best possible SOS signal-to-noise ratio after $n$ repeats is given by

$$SNR_{SOS} = \frac{\sqrt{n}\, S_{SOS}}{\sqrt{\delta^2 S_{SOS}}} = \frac{|\langle k_0\rangle - \langle k_1\rangle|}{\sqrt{(\delta^2 k_0 + \delta^2 k_1)}} \sqrt{n}, \tag{S8}$$

where we dropped the time arguments for simplicity, and used the fact that both the signal and variance grow linearly with the number of repeats. We finally obtain for the SOS sensitivity

$$\eta_{SOS} = \frac{\sqrt{\delta^2 S_{SOS}}}{\sqrt{n}\, S_{SOS}} \sqrt{n\, t_{SOS}} = \frac{\sqrt{t_{SOS}(\langle k_0\rangle + \langle k_1\rangle)}}{|\langle k_0\rangle - \langle k_1\rangle|}, \tag{S9}$$

where we used $\delta^2 k_0 = \langle k_0\rangle$ and $\delta^2 k_1 = \langle k_1\rangle$, valid for a Poisson distribution. The expression in the main text follows by noting that $\langle k_1\rangle = (1 - \mu)\langle k_0\rangle$ with $0 < \mu < 1$.

*ii-SCC sensitivity*
Now, we turn our attention to the action of the SCC scheme with local optical readout, which can be modeled via the operator

$$M_{SCC} = k_0'|0\rangle\langle 0| + k_1'|1\rangle\langle 1|, \tag{S10}$$

where $k_0'$ and $k_1'$ are stochastic variables indicating the spin-dependent photon counts after the SCC pulse. Following the reasoning above, we find

$$S_{SCC}(t_e, t_r) = |u_1|^2 |\langle k_0'\rangle(t_r) - \langle k_1'\rangle(t_r)|, \tag{S11}$$

and

$$\delta^2 S_{SCC} = |u_0|^2 \delta^2 k_0' + |u_1|^2 \delta^2 k_1' + \delta^2 k_0' \geq \delta^2 k_0' + \delta^2 k_1'. \tag{S12}$$

Assuming as before $|u_1|^2 = 1$, we then write

$$SNR_{\text{SCC}} = \frac{\sqrt{n}\, S_{\text{SCC}}}{\Delta S_{\text{SCC}}} = \frac{|\langle k_0' \rangle - \langle k_1' \rangle|}{\sqrt{(\delta^2 k_0' + \delta^2 k_1')}} \sqrt{n}\,. \qquad (S13)$$

To find an explicit expression for $\Delta S_{\text{SCC}}$, we assume that

- Charge conversion of the qubit is limited exclusively to the SCC pulse, i.e., no charge change takes place during readout;
- Only one charge state of the qubit emits fluorescence (e.g., we detect photons exclusively from NV$^-$)

Under these conditions, we model SCC as a multi-step stochastic process, namely, we express the spin-dependent photon counts as

$$k_j' = \bigl(p(1 - q_j) + (1 - p)r\bigr)k\,, \qquad (S14)$$

where $p$, $q_j$, $r$, and $k$ are stochastic variables, and $j = \{0,1\}$ denotes the two qubit spin projections. More specifically,

- $q_j = \{0,1\}$ is a Boolean stochastic variable associated with the spin-to-charge conversion probability of the qubit (e.g., the probability of transforming NV$^-$ into NV$^0$). We describe the dynamics via the binomial distribution $\mathcal{B}_{q_j}$ with mean value $\langle q_j \rangle$.
- $p = \{0,1\}$ is a Boolean stochastic variable describing the probability of finding the qubit in the 'desired' charge state prior to SCC ionization (e.g., the negatively charged state for an NV). This variable follows the binomial probability distribution $\mathcal{B}_p$ with mean value $\langle p \rangle$.
- $r = \{0,1\}$ is a Boolean stochastic variable associated with the probability of transforming the qubit charge state from the spin-inactive to its spin active state during the SCC pulse (e.g., transforming NV$^0$ into NV$^-$). It is governed by the binomial distribution $\mathcal{B}_r$ with mean $\langle r \rangle$.
- $k = \{0,1,2,3\ldots\}$ is a stochastic variable describing the emitted photons during readout and is governed by a Poisson distribution $\mathcal{P}_k(\langle k \rangle)$ with mean value $\langle k \rangle$. Note that because we assume the qubit does not change its charge state during readout, $\langle k \rangle$ is simply given by the product between the qubit emission rate and the readout time.

With these definitions, the first contribution in Eq. (S14) can be interpreted as the number of photons emitted by a spin qubit that started and remained with the spin-active charge state (e.g., NV$^-$) throughout the SCC pulse; on the other hand, the second term stems from qubit emission that was 'erroneously' brought back into the spin-active charge state during the SCC pulse (e.g., NV$^0$ transforming into NV$^-$).

Assuming for simplicity $|u_1|^2 = 1$, we can now rewrite the SCC signal as

$$S_{\text{SCC}} = \langle p \rangle |\langle q_0 \rangle - \langle q_1 \rangle| \langle k \rangle\,. \qquad (S15)$$

More importantly, we recast the variance as

$$\delta^2 S_{\text{SCC}} = \sum_{j=0}^{1} \bigl(\langle v_j \rangle^2\, \delta^2 k + \langle k \rangle^2\, \delta^2 v_j + \delta^2 k \cdot \delta^2 v_j\bigr) = \sum_{j=0}^{1}\bigl(\langle k \rangle \langle v_j \rangle^2 + (\langle k \rangle^2 + \langle k \rangle)\, \delta^2 v_j\bigr)\,, \qquad (S16)$$

where we define the combined stochastic variables $v_j \equiv p(1 - q_j) + (1 - p)r$, $j = \{0,1\}$ and use the fact that, for a Poisson distribution, $\delta^2 k = \langle k \rangle$. Assuming all variables are independent, we attain the final expression for $\delta^2 S_{\text{SCC}}$ simply by propagating the variance through the sum and products in $s_j$, and by using the known expression for the variance of binomial distributions. After some algebra, we find

$$\delta^2 v_j = \bigl(1 - \langle q_j \rangle\bigr)^2 \langle p \rangle(1 - \langle p \rangle) + \langle p \rangle^2 \langle q_j \rangle(1 - \langle q_j \rangle) + \langle r \rangle^2 \langle p \rangle(1 - \langle p \rangle) + (1 - \langle p \rangle)^2 \langle r \rangle(1 - \langle r \rangle)$$
$$+ \langle p \rangle(1 - \langle p \rangle)\bigl(\langle q_j \rangle(1 - \langle q_j \rangle) + \langle r \rangle(1 - \langle r \rangle)\bigr)\,. \qquad (S17)$$

We can therefore write the SCC sensitivity as

$$\eta_{\text{SCC}} = \frac{\sqrt{t_{\text{SCC}}}\left((\langle k\rangle^2 + \langle k\rangle)(\delta^2 v_0 + \delta^2 v_1) + \langle k\rangle(\langle v_0\rangle^2 + \langle v_1\rangle^2)\right)^{1/2}}{\langle p\rangle|\langle q_0\rangle - \langle q_1\rangle|\langle k\rangle}. \tag{S18}$$

where $t_{\text{SCC}} \equiv t_i' + t_r' + t_{\text{scc}} + t_e$, with each term representing the qubit charge initialization and readout times, the SCC pulse duration, and the spin evolution time, respectively.

*iii-AID sensitivity*

Since the dynamics at play during AID is nearly identical to that in SCC, the above formalism can be easily extended once we note that photon collection stems here from an 'activated' ancilla defect, not the qubit itself. With this in mind, we define the AID measurement operator as

$$M_{\text{AID}} = k_0''|0\rangle\langle 0| + k_1''|1\rangle\langle 1|, \tag{S19}$$

where $k_j''$, $j = \{0,1\}$ are stochastic variables describing the number of photons collected from the ancilla defect for a qubit initially in one spin state or the other. To model the underlying photon generation, diffusion, and capture process, we write $k_j''$ as

$$k_j'' = \lambda(pq_j + w)k_a, \tag{S20}$$

where $k_a$ is the number of photons emitted by an activated ancilla defect, $w$ is the number of carriers produced by the ionization of background defects at the site of illumination, and $\lambda$ is a stochastic variable associated with the carrier capture probability by the ancilla trap ensemble, here viewed as a measure of the 'carrier collection efficiency' (in general a function of the spatial distribution of ancilla defects around the qubit).

By analogy to Eqs. (S15) and (S16) and assuming $|u_1|^2 = 1$, we conclude that

$$S_{\text{AID}} = \langle\lambda\rangle\langle p\rangle|\langle q_0\rangle - \langle q_1\rangle|\langle k_a\rangle, \tag{S21}$$

and

$$\delta^2 S_{\text{AID}} = \sum_{j=0}^{1} \delta^2(\lambda v_j k_a) = \langle\lambda\rangle \sum_{j=0}^{1}\left[\delta^2(v_j k_a) + \langle k_a\rangle^2\langle v_j\rangle^2(1-\langle\lambda\rangle)\right], \tag{S22}$$

where we defined $v_j \equiv pq_j + w$ and assumed $\lambda$ obeys a binomial distribution. By the same token, we write

$$\delta^2(v_j k_a) = \langle k_a\rangle(\langle v_j\rangle^2 + \delta^2 v_j) + \langle k_a\rangle^2 \delta^2 v_j, \tag{S23}$$

and

$$\delta^2 v_j = \langle q_j\rangle^2\langle p\rangle(1-\langle p\rangle) + \langle p\rangle^2\langle q_j\rangle(1-\langle q_j\rangle) + \langle p\rangle(1-\langle p\rangle)\langle q_j\rangle(1-\langle q_j\rangle) + \langle w\rangle, \tag{S24}$$

where we assume $w$ obeys a Poisson distribution. Combining Eqs. (S21) through (S24), we finally write the AID sensitivity after $n$ repeats as

$$\eta_{\text{AID}} = \frac{\sqrt{t_{\text{AID}}}\left(\sum_{j=0}^{1}[\delta^2(v_j k_a) + \langle k_a\rangle^2\langle v_j\rangle^2(1-\langle\lambda\rangle)]\right)^{1/2}}{\sqrt{\lambda}\,\langle k_a\rangle\,\langle p\rangle\,|\langle q_0\rangle - \langle q_1\rangle|} \tag{S25}$$

where $t_{\text{AID}} \equiv t_{\text{scc}} + t_e + \frac{t_{ia}' + t_{ra}'}{n}$ is the average time per duration and $t_{ia}'$, $t_{ra}'$ are the ancilla ensemble charge initialization and readout times, respectively.

*iv-The role of background impurities*

Unlike NVs — transforming from NV⁻ to NV⁰ upon hole capture but insensitive to electrons[1] — nitrogen impurities (and possibly other defects) present in the sample can dynamically reconvert from one charge state to the other when both electrons and holes are present. As discussed in the main text, this process gradually reduces the number of carriers able to activate outer lying NVs in the ancilla ensemble, hence leading to a progressively slower signal growth. To gauge the impact this effect has on the AID sensitivity, we first lay out and solve the set of master equations describing the photo-generation, diffusion, and capture of carriers in the presence of continuous optical excitation. For simplicity, we consider a diamond sample comprising only NV centers and N impurities with total volume concentration $Q$ and $P$, respectively; the relevant quantities describing the charge dynamics are the NV⁻ density $Q_-(\mathbf{r},t)$, the density of neutral nitrogen $P_0(\mathbf{r},t)$, and the densities of holes $p(\mathbf{r},t)$ and electrons $n(\mathbf{r},t)$, all functions of position $\mathbf{r}$ and time $t$. The complete set of equations is[1,2]

$$\frac{\partial Q_-}{\partial t} = (\vartheta_0 + \kappa_n n)Q - (\vartheta_0 + \vartheta_- + \kappa_n n + \kappa_p p)Q_- + \Omega[Q(P - P_+) - Q_- P],$$

$$\frac{\partial P_+}{\partial t} = (\vartheta_N + \gamma_p p)P - (\vartheta_N + \gamma_n n + \gamma_p p)P_+ + \Omega[Q(P - P_+) - Q_- P],$$

$$\frac{\partial n}{\partial t} = D_n \nabla^2 n + \vartheta_- Q_- + \vartheta_N (P - P_+) - \kappa_n n(Q - Q_-) - \gamma_n n P_+,$$

$$\frac{\partial p}{\partial t} = D_p \nabla^2 p + \vartheta_0 (Q - Q_-) - \kappa_p p Q_- - \gamma_p p(P - P_+),\quad\quad (S26)$$

where $\vartheta_0$ and $\vartheta_-$ denote the photo-induced NV⁰→NV⁻ and NV⁻→NV⁰ conversion rates, respectively. These rates depend on the intensity of illumination $I(\mathbf{r})$ and wavelength $\Lambda$; $\kappa_p$ and $\kappa_n$ denote the NV⁻ hole capture rate and the NV⁰ electron capture rate, respectively. The nitrogen ionization rate[9] is given by $\vartheta_N$ (= $k_N(I(\mathbf{r}),\Lambda)$); neutral nitrogen (N⁰) can capture holes at a rate $\gamma_p$ and ionized nitrogen (N⁺) can capture electrons at a rate $\gamma_n$. In addition, coefficients $D_n$ and $D_p$ are the electron and hole diffusion constants at room temperature, respectively, and $\nabla^2$ denotes the Laplace operator in two-dimensions. The previous set of equations is based on three assumptions that make our problem tractable. First, global charge conservation is ensured by imposing

$$\int_V e[(P - P_0) - Q_- + p - n]dr^3 = 0 \quad\quad (S27)$$

where $e$ is the fundamental charge and the integral is over the diamond volume $V$. Second, we neglect the electric forces arising from non-uniform charge density distributions, which means that the carrier dynamics is entirely ruled by free diffusion and trapping (for a detailed discussion on this assumption see Supplemental Ref. [2]). Third, cylindrical symmetry is assumed, which reduces the problem to two dimensions.

We solve the set of equations (S26) using Matlab Partial Derivative Equation solver for parabolic equations with cylindrical symmetry. The following scenario is examined: every AID cycle contains a spin-to-charge conversion laser pulse that is comprised of combined or consecutive 532 nm and 632 nm illumination. The duration of this illumination is set to be 80 ns per cycle corresponding to the single NV spin-to-charge conversion pulse[3]. We assume that the cumulative effect of repeating this illumination pulse with a given duty cycle (as in the present experiments) is equivalent to the continuous wave illumination of the duration that corresponds to 80 ns × $n$, where $n$ is the total number of repeats. Ionization of N⁰ and recharging of NV generates holes and electrons at the illumination point, which then diffuse and get captured by charge traps outside the illumination spot. We sample a circular area with 250 μm radius and assume that 70% of the NV population in this area has been prepared into the NV⁻ ancilla state, 'activated' into the neutral charge state by capturing holes, i.e., NV⁻ + p → NV⁰. The activated ancilla trap state NV⁰ is immune to charge carriers ($\sigma_{NVe}=0$). Under these conditions, nitrogen impurities act as background traps capturing holes while in the neutral state ($\sigma_{Np} = 1.4 \cdot 10^{-8}$ μm²) and electrons while in the positive state ($\sigma_{Np} = 3.1 \cdot 10^{-6}$ μm²). The quantity of interest for our calculations is the number of activated ancilla NVs as a function of illumination time or, equivalently, the number of repeats. We obtain the number of activated ancilla NVs from the 2D cylindrical distribution of concentration assuming the thickness of the plane to be 1 μm. In order to quantify the effect of the background traps we perform calculations for different capture cross sections of these traps $\sigma = 0$; $0.01 \cdot \sigma_{Np}$; $0.1 \cdot \sigma_{Np}$; $0.5 \cdot \sigma_{Np}$; $\sigma_{Np}$ and $2 \cdot \sigma_{Np}$. The list of parameter values used in Eq. (S19) is summarized in Table S1 along with corresponding references. NV⁻ (N⁰) hole and NV⁰ (N⁺) electron capture rates are determined as $\kappa_p = \sigma_{NVp} * v$ ($\gamma_p = \sigma_{Np} * v$) and $\kappa_n = \sigma_{NVn} * v$ ($\gamma_n = \sigma_{Nn} * v$), where $\sigma_{NVp}$ ($\sigma_{Np}$) and $\sigma_{NVn}$ ($\sigma_{Nn}$) are NV⁻ (N⁰) hole and NV⁰ (N⁺) electron capture cross sections, respectively, and v denotes the electron thermal velocity. The effective laser power is chosen to be 2.22 mW so that the number of activated ancilla traps per experimental cycles matches the experimental value of 0.8 charges generated on average per cycle corresponding to the experimental single NV spin-to-charge efficiency[3]. This is ensured by monitoring the number of activated ancilla traps in the case when the background traps are set to have hole capture cross section equal to 0. This scenario implies that all the generated holes are being captured by ancilla traps with 100 % efficiency, and it corresponds to the linear dependence in Fig. S1c below.[4,5]

| | | | |
|---|---|---|---|
| $P$ | Substitutional Nitrogen density | 1 ppm | |
| $Q$ | Nitrogen Vacancy density | 0.01 ppm | |
| $Q_-$ | Initial NV⁻ density | 0.007 ppm | |
| $P_0 = P - Q_-$ | Initial N⁰ density | 0.993 ppm | |
| $\sigma_{Nn}$ | N⁺ electron capture cross section | $3.1 \cdot 10^{-6}$ μm² | [4] |
| $\sigma_{Np}$ | N⁰ hole capture cross section | $1.4 \cdot 10^{-8}$ μm² | [4] |
| $\sigma_{NVp}$ | NV⁻ hole capture cross section | $9 \cdot 10^{-8}$ μm² | [1] |
| $\sigma_{NVe}$ | NV⁰ electron capture cross section | 0 | [1] |
| $I$ | 532 nm illumination power | 2.22 mW | |
| $I_0$ | Reference 532 nm illumination power | 1 μW | |
| $s$ | Variance of 532 nm laser beam Gaussian power spatial distribution | 1 μm | |
| $\vartheta_N(r)$ | N⁰ photoionization rate under 532 nm illumination | $15 \cdot \frac{I}{I_0} \cdot \exp\left(-\frac{r^2}{s^2}\right)$ Hz | [1] |
| $\vartheta_0(r)$ | NV⁰→NV⁻ recombination rate under 532 nm illumination | $0.0046 \cdot \left(\frac{I}{I_0}\right)^2 \cdot \exp\left(-\frac{r^2}{s^2}\right)$ Hz | [1] |
| $\vartheta_-(r)$ | NV⁻→NV⁰ photoionization rate under 532 nm illumination | $0.0107 \cdot \left(\frac{I}{I_0}\right)^2 \cdot \exp\left(-\frac{r^2}{s^2}\right)$ Hz | [1] |
| $\mu_n$ | Electron mobility in diamond | $2.4 \cdot 10^{11}$ μm²/(V·s) | [4,5] |
| $\mu_p$ | Hole mobility in diamond | $2.1 \cdot 10^{11}$ μm²/(V·s) | [4,5] |
| $D_n = \frac{\mu_n k_B T}{e}$ | Electron diffusion coefficient in diamond | $6.1 \cdot 10^9$ μm²/s | |
| $D_p = \frac{\mu_p k_B T}{e}$ | Hole diffusion coefficient in diamond | $5.3 \cdot 10^9$ μm²/s | |

**Table S1**. List of parameters used in calculation of trapping efficiency by the ancilla traps according to Eq. (S19). Listed values are for ambient conditions T=293 K; $k_B = 1.38 \cdot 10^{-23} \frac{J}{K}$ stands for Boltzmann constant, and $e = 1.6 \cdot 10^{-19}$ C is the elementary charge. Last column provides references for the values taken from previous experimental studies.

**Supplementary Note 3 | Monte Carlo simulations**

In this section, we describe the Monte Carlo simulations used to model the ODMR spectra and sensitivities of the SOS and AID protocols. We made use of global charge conservation, neglected space charge effects, and assumed that the only defects present are NVs and substitutional nitrogen. We simulated spectra and sensitivity performance for AID-ODMR by sampling statistical distributions associated with the carrier generation and capture processes (binomial distributions) and the photo-emission process from defects (Poissonian distributions). These simulations used built-in statistical sampling functions in the Matlab programming platform.

*i-SOS ODMR*

We simulate SOS data by sampling and summing photon number realization $k_i$ extracted from a Poissonian distribution $\mathcal{P}(\langle k(f) \rangle)$ with a mean number of emitted photons $\langle k \rangle(f)$, in general, a function of the MW frequency $f$. Far off resonance, $\langle k \rangle(f) = \langle k_0 \rangle = 0.075$ is the average number of photons collected from a single NV$^-$ (initially in the $m_S = 0$ state) in a 300-ns counting window under saturation illumination with 532 nm light[6]. To model the action of MW on the NV spin, $\langle k \rangle(f)$ is modulated by a Lorentzian function $L(f) = 1/[1 + x^2(f)]$ where $x(f) = 2(f - f_\Delta)/\Delta f$, $f_\Delta = 2.87$ GHz (the NV$^-$ crystal-field resonance), and the FWHM of the function is $\Delta f = 7$ MHz. As the MW frequency is swept, the Bloch vector representing the spin state of the NV$^-$ rotates from the $m_S = 0$ state (bright, $\langle k_0 \rangle$) to $m_S = 1$ (dark, $\langle k_1 \rangle = (1 - \mathbb{C}_{SOS}) \cdot \langle k_0 \rangle$). We take the contrast between the average number of photons collected in the bright and dark spin states to be $\mathbb{C}_{SOS}=0.3$, giving[7] $\langle k_1 \rangle = 0.51$. Each point in the MW sweep is sampled $n = 10^4$ times, and the photons accumulated from all realizations at a given frequency are summed to generate a plot of the spectrum. The SNR of the plot is calculated from two data sets with the MW on and off resonance. The signal is taken as the difference in the average number of photons collected in both states ($S_{SOS} = n|\langle k_0 \rangle - \langle k_1 \rangle|$), and subsequently determine the variance $n \delta^2 k_j = n(\langle k_j^2 \rangle - \langle k_j \rangle^2)$, $j = \{0,1\}$ of each generated data set, giving $SNR_{SOS} = \frac{\sqrt{n} |\langle k_0 \rangle - \langle k_1 \rangle|}{(\delta^2 k_0 + \delta^2 k_1)^{1/2}}$.

*ii-AID ODMR*

In our AID ODMR simulations, electrons generated by the SCC cycle ($\langle q_0 \rangle = 0.8, \langle q_1 \rangle = 0.5$)[3] diffuse away from a single qubit NV center and are captured by ancilla defects, converting them to their bright state, i.e. "activating" them. While the process $NV^- + h^+ \rightarrow NV^0$ is experimentally realized in NV-AID (i.e. conversion of NV ancillae to their dark state), we make the simplifying assumption that carriers originating from the qubit NV are captured by the ancilla and convert them to the bright state. This assumption is not a fundamental difference in the physics of the simulation, merely causing a decrease in collected photons when the MW is on resonance rather than an increase and hence making modeling more convenient. We take the initial charge state of the qubit NV to be NV$^-$ with $\langle p \rangle = 1$, and we set $\langle r \rangle = 0$ (see Supplementary Note I).

Charge dynamics are treated as a two-step process: First, the SCC protocol is modeled by sampling a binomial distribution $\mathcal{B}_q(\langle q \rangle(f))$ where $\langle q \rangle(f)$ is the probability of success (i.e., ionizing a charge in the bright spin state), here undergoing the same Lorentzian dependence $L(f)$ with $\mathbb{C}_{AID} = 0.36$. Each successful realization triggers a second random sampling extracted from a binomial distribution $\mathcal{B}_\lambda(\langle \lambda \rangle(i))$, with mean $\langle \lambda \rangle(i)$, in general, a function of the index $i = 1 \ldots n$ in a run with $n$ realizations (see below).

Readout of the ancillary defects is modeled with a wide-field sensor (requiring no raster scanning, e.g., via a CCD) and 594 nm illumination at 1 $\mu$W/$\mu$m$^2$ for 10 ms readout time giving $\langle k_a \rangle = 22$ photons per readout for a single defect[6]. Under these conditions, the ionization rate of the NV$^-$ is approximately[6] 24 Hz and the ancillae remain in their bright charge state with ~80% probability, so we disregard ionization/recombination dynamics during this time. Photon statistics are generated by sampling a Poissonian distribution $\mathcal{P}_{k_a}(\langle k_a \rangle)$ for each activated ancilla. The SNRs for the AID spectra are calculated in the same manner as in SOS.

*iii-Comparison between AID and SOS sensitivities*

Starting with the geometry in Fig. S1a, Fig. S1b shows the results from a numerical comparison with SOS upon application of the protocol in Fig. 3A for the case of a single NV qubit surrounded by NV ancillae. Using $\langle k_0 \rangle = 0.075$ ($\langle k_1 \rangle = 0.05$) to denote the number of photons detected per optical spin readout for $m_S = 0$ ($m_S = \pm 1$), we set the number of repeats $n = 10^4$ so as to obtain a moderate signal-to-noise ratio (SNR) upon SOS detection (upper left plot in Fig. S1b). Wide field excitation together with charge coupled device (CCD) detection can (hypothetically) bring down the combined charge initialization and readout time $t'_{ia} + t'_{ra}$ during AID to only a small fraction of the typical averaging times $n(t_{scc} + t_e) \gtrsim 1$ s. Since $\langle k_a \rangle \sim 260 \langle k_0 \rangle$ for an NV under optimal wavelength and laser intensity[8], AID can potentially yield greater SNR even in the least favorable regime of short spin evolution times (we use $t_e = 15$ μs in Fig. S1b).

More rigorously, we write the SOS sensitivity as

$$\eta_{SOS} \sim \frac{\sqrt{2}}{\mu \sqrt{\langle k_0 \rangle}} (t_i + t_r + t_e)^{1/2}, \tag{S28}$$

where $\mu$ satisfies $\langle k_1 \rangle = (1 - \mu)\langle k_0 \rangle$, and $t_i + t_r$ is the combined optical initialization and readout time of the qubit spin. Assuming the regime of Eq. (2) and for $\langle \lambda \rangle \sim 1$, the formal criterion for $\eta_{AID} < \eta_{SOS}$ can then be cast as

$$\mu^2 \langle k_0 \rangle \left(1 + \frac{t'_{ia} + t'_{ra}}{nt_c}\right) < \frac{|\langle q_0 \rangle - \langle q_1 \rangle|}{\langle q_0 \rangle(1 - \langle q_0 \rangle) + \langle q_1 \rangle(1 - \langle q_1 \rangle)}, \tag{S29}$$

where $t_c = t_i + t_r + t_e \approx t_{scc} + t_e$ is the protocol time per cycle (similar for both AID and SOS.

Our present experimental conditions are far from this ideal regime, partly because the time intervals $t'_{ia}$, $t'_{ra}$ required for point-by-point charge initialization and readout as implemented herein are intrinsically long. Further, background traps coexisting with the ancillae (most notably, nitrogen impurities) can potentially trap spin-encoded carriers and dynamically convert from and to a starting charge state upon successive capture of electrons and holes diffusing from the qubit (see below). Both mechanisms effectively reduce the number of signal-carrying charges able to activate additional ancilla defects, hence leading to a non-linear, gradually slower signal growth (as observed in Figs. 2b and 3e in the main text) and thus to a reduced SNR (lower right plot in Fig. S1b). Note that optical excitation of background defects at the point of illumination can also produce information-less, background carriers whose capture by the ancilla traps negatively impacts the AID SNR (lower left plot in Fig. S1b, see below). In the limit dominated by poor ancilla activation efficiency (i.e., $\langle \lambda \rangle \ll 1$) and multiple background carriers (i.e., $\langle w \rangle \gg \langle p \rangle \langle q_j \rangle$, $j = \{0,1\}$), Eq. (1) yields

$$\eta_{AID} \sim \frac{\sqrt{2 t_{AID}} \langle w \rangle}{\sqrt{\langle \lambda \rangle} \langle p \rangle |\langle q_0 \rangle - \langle q_1 \rangle|}, \tag{S30}$$

where the advantages of integrated charge readout have vanished (the case for our present diamond sample). Interestingly, the sensitivity continues to worsen as $\langle \lambda \rangle$ decreases, even if the observed contrast — here corresponding to $\mathbb{C}_{AID} \sim \langle p \rangle |\langle q_0 \rangle - \langle q_1 \rangle|/\langle w \rangle \sim 4\%$ — remains non-zero.

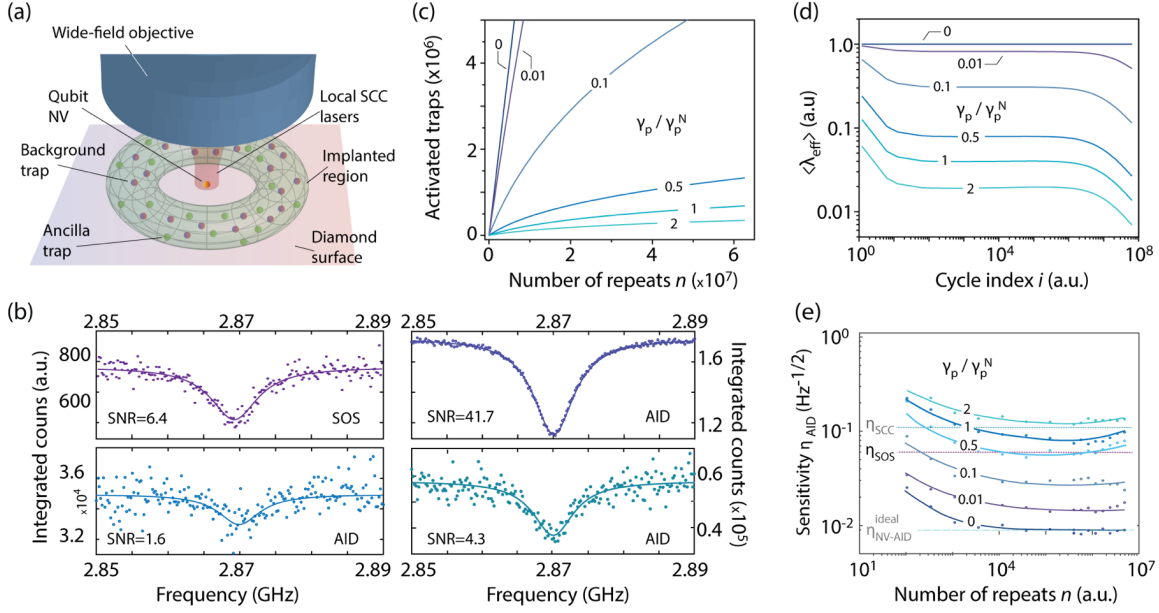

**FIG. S1. Quantifying detection sensitivity.** (a) Model AID geometry. Photo-generated carriers diffuse from a central NV qubit into an ensemble of surrounding traps comprising ancilla defects and background, non-fluorescent traps. We assume charge initialization and readout of the entire ensemble via wide field illumination on a time $t'_{\text{ia}} + t'_{\text{ra}} = 10$ ms. (b) (Upper left) Simulated SOS magnetic resonance spectrum from a single NV⁻ spin qubit using $\langle k_0 \rangle = 0.075$. (Upper right) Simulated AID magnetic resonance spectrum assuming $\lambda = 1$ and $\langle k_a \rangle = 22$ in the absence of background traps. (Lower right) Same as before but in the presence of nitrogen background traps within the capture volume. (Lower left) Same as before but with the contribution from background traps and background carriers ($\langle w \rangle = 4$). In all cases, $n = 10^4$ and $t_e = 15$ μs; see text for further details. (c) Number of activated ancilla NVs as a function of $n$ in the presence of background traps with relative hole-capture rate $\gamma_p/\gamma_p^N$. In all cases, the relative concentrations of ancilla and background traps satisfy 1:100, and we assume equal initial number of ancilla (background) traps in the negative (positive) charge state. (d) Effective ancilla activation probability $\lambda_{\text{eff}}$ as a function of the cycle index $i = 1 \ldots n$. (e) Calculated sensitivity $\eta_{\text{AID}}$ as a function of $n$; the horizontal dashed (dotted) line indicates the value for $\eta_{\text{AID}}$ ($\eta_{\text{SOS}}$) calculated in (c), and we include $\eta_{\text{SCC}}$ for the same conditions. In (b) through (d), data points indicate the results of a Monte Carlo simulation and solid traces represent analytical or numerical solutions; in (e) solid lines are guides to the eye. For simplicity, we assume in our simulations that the outer radius of the implanted region is infinite.

To more thoroughly quantitate the interplay between background traps and sensitivity, we first calculate in Fig. S1c the time evolution of the charge pattern in Fig. S1a assuming the ancilla traps (NVs in this example) coexist with 100-fold more abundant $N$ impurities (as found in our sample). The plotted numerical response emerges from solving the set of coupled master equations governing the generation, diffusion, and trapping of photo-induced carriers (see below), which we then use to derive the effective trapping efficiency $\langle \lambda_{\text{eff}} \rangle(n)$ during the $n$-th repeat (Fig. S1d). We subsequently adapt our Monte Carlo simulations in a way that takes into account the time-dependent ancilla trap activation, and numerically determine $\eta_{\text{AID}}$ as a function of the integrated number of repeats (Fig. S1e). We find, in general, a complex function of $n$, with an optimum that depends on the carrier capture selectivity of the background traps. In particular, $\eta_{\text{AID}}$ reaches virtually ideal levels if the hole capture cross section $\gamma_p$ of the background traps — here referenced to $\gamma_p^N$, the value for nitrogen[9] — is brought to zero, though we find that even a moderate hole-capture probability can have a damaging effect. More realistically, we anticipate a close-to-ideal response if the relative concentration of background traps such as nitrogen is reduced to values comparable to the NVs (see below).

*iv-Effect of background defects*

As mentioned above, carriers generated during SCC may be trapped not only by ancillary defects in the readout region, but also by background defects that emit no photoluminescence (PL). As a result, the number of carriers that are trapped by ancilla defects does not grow linearly with the number of repetitions of the SCC cycle (Fig. S1c). For a given type and concentration of background defects, the capture probability $\langle \lambda \rangle$ becomes a function of the number $i = 1 \ldots n$ indexing each run in an experiment totaling $n$ repeats (Fig. S1d). Figure S1b shows the NV magnetic resonance spectrum generated with background charge capture included; in this particular case, we assume nitrogen background defects with relative concentration 100-fold greater than the NV. More generally, we expose the effect of background defects either by varying the type of defect present — here attained by changing the carrier cross section $\gamma_p = \varepsilon \gamma_p^N$ relative to $\gamma_p^N$, the hole capture cross section for nitrogen[10], Figs. S1d and S1c — or, for the common case of nitrogen, by varying the background defect concentration.

Figure S1b (lower left) shows a spectrum including the effect of background defects being ionized and producing carriers that subsequently activate ancilla defects but convey no spin information. The number $w$ of carriers produced by ionization of background defects during the SCC cycle is generated in a manner similar to the single NV$^-$ case, except that there is no impact of the MW excitation on the ionization rate (we use $\langle q \rangle_w = 0.8$ per background defect within the illumination spot). The latter results in an increase of activated ancillary defects and thus an increase in PL at all MW frequencies. Since this signal is spin independent, it reduces $\mathbb{C}_{AID}$ and thereby reduces the SNR.

Fig. S1e shows sensitivity curves generated by numerically determining the SNR of the simulated AID ODMR protocol and subsequently taking the ratio of the total time of the experimental sequence and the SNR, i.e., $\eta_{AID} = \frac{\sqrt{n\, t_{AID}}}{SNR_{AID}}$. We use $t_{AID} = t_e + t_{scc} + (t'_{ra} + t'_{ia})/n$ with charge initialization and readout times extracted from Ref. [3] assuming wide-field optical excitation. To highlight the effect of the trapping background on $\eta_{AID}$, each curve is simulated with a different trapping coefficient $\langle \lambda \rangle_\varepsilon(i)$, each a function of count index $i$ as derived in Fig. S1d for different relative trapping cross sections $\varepsilon = \gamma_p/\gamma_p^N = \{0, 0.01, 0.5, 1, 2\}$, as extracted from Fig. S1d.

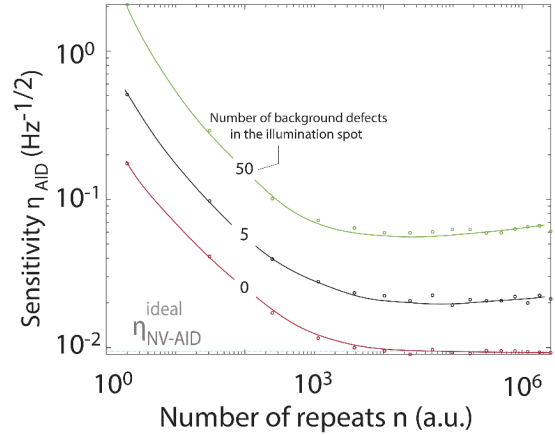

**FIG. S2**. **The impact of background carriers.** Monte Carlo simulation of AID sensitivity vs. the number of experimental repeats, including the effects of background charge carriers being generated during the SCC cycle and subsequently activating ancillary defects. Each curve is simulated with a different number of background defects in the illumination spot (0, 5 and 50), and the probability that a background defect is ionized during a SCC cycle is the same as the qubit NV (0.8).

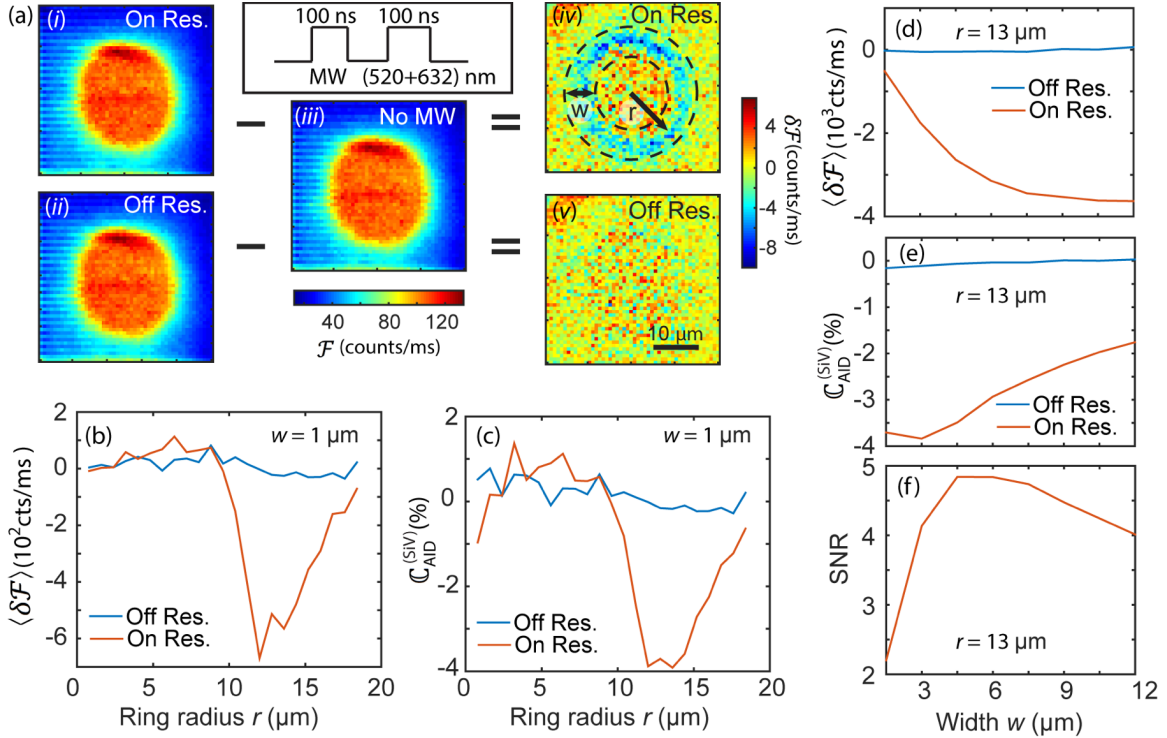

**FIG. S3. NV-spin to SiV-charge state conversion in AID.** (a) (*i*) through (*iii*) are the 632 nm, 1.5 mW (40×40 μm$^2$) scans after 1 second of spin-to-charge conversion with on-resonance MW (2.87 GHz), with off-resonance MW (2.84 GHz), and without MW, respectively. (*iv*) is the difference plot between (*i*) and (*iii*), and (*v*) is the difference plot between (*ii*) and (*iii*). To calculate the qubit spin signal, we integrate the number of counts along the ring (width $w$ and radius $r$, see (*iv*)). (b) Difference between the integrated counts for on-resonance (off-resonance) MW excitation and no MW as a function of the radius $r$ for a fixed width of 1 μm. (c) Contrast relative to the no-MW data. (d, e, f) Respectively, differential counts, contrast and SNR for a fixed radius of $r = 13$ μm and a variable ring width $w$.

In Fig. S2, we show a simulation of the sensitivity of the AID-ODMR technique as a function of experimental repetitions, but with the contribution of background carrier ionization included. We assume each background trap in the beam spot has an ionization probability that is the same as the qubit NV, and that each generated background carrier has the same probability of capture as a carrier that encodes spin information (in this case, $\varepsilon = 1$). This simulation emphasizes the capacity of the AID technique to improve if technical conditions are more precisely engineered (e.g. a diamond with fewer background defects).

**Supplementary Note 4 | Extracting the qubit spin signal from SiV$^-$- and NV$^-$-based AID readout**

In this section, we provide additional details on how we derive the qubit spin signal from the charge patterns in their vicinity. For the case of activated SiVs, the differential pattern that emerges from subtracting the on- and off-resonance MW excitation patterns takes the form of a dark ring concentric with the point of illumination. To quantitatively characterize the spin signal extracted from these differential patterns, we use $r$ and $w$, respectively denoting the average distance to the point of illumination and ring width. Fig. S3 shows a summary of this characterization as a function of $r$ and $w$, allowing us to find the conditions for optimal SNR: For example, in Figs. S3b and S3c we plot the (azimuth) integrated differential fluorescence and contrast using a minimal ring width. We find both the differential fluorescence and signal contrast peak in the region around $r = 13$ μm. Choosing this value as the optimum average radius, we gradually increase the ring width $w$ (Figs. S3d and S3e). We find that although the differential fluorescence $\delta \mathcal{F}$ grows monotonously (Fig. S3d), the contrast $\mathbb{C}_{AID}^{(SiV)}$ reaches a maximum at $w = 3$ μm (Fig. S3e). On the other hand, a maximum SNR is attained using $w = 5$ μm, the value we use throughout the main text.

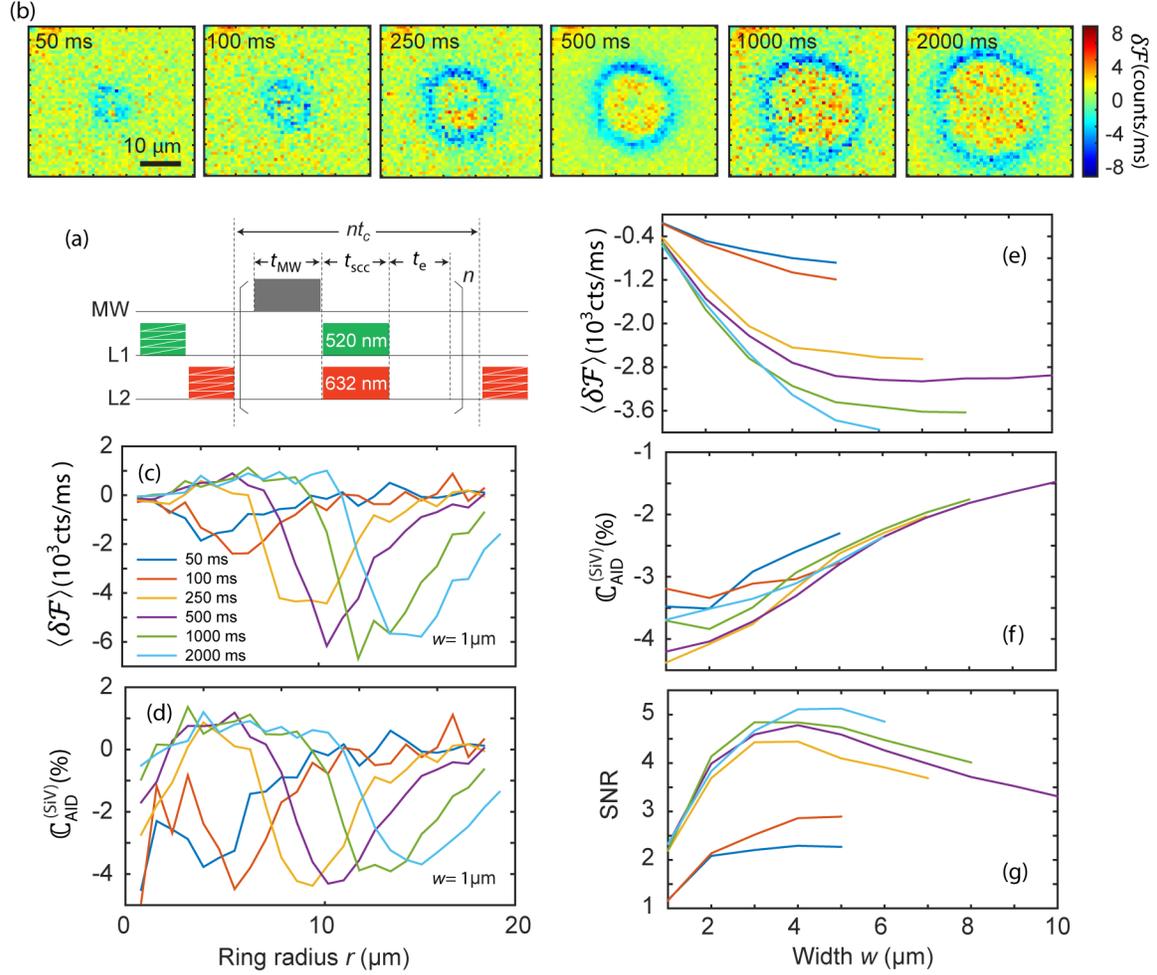

**FIG. S4. SiV-AID signal time buildup.** (a) Pulse sequence. The green laser power is 3 mW both during the initialization scan and SCC; the red laser power is 2 mW during initialization, 21 mW during SCC, and 1 mW during readout. The total ancilla initialization (readout) time is $t'_{ia} = 40$ s ($t'_{ra} = 5$ s); scanning extends over an area of ($40 \times 40$ μm$^2$) centered at the point of qubit illumination, with each pixel size corresponding to 0.64 μm$^2$. (b) Differential SiV$^-$ photoluminescence images upon application of the protocol in (a) with and without on-resonance MW at 2.87 GHz. The time in the upper right corner of each image is $nt_c = n(t_{scc} + t_{MW} + t_e)$, where $t_{scc} = t_{MW} = 100$ ns and $t_e = 1$ μs is a wait time between successive cycles. (c, d) Signal counts and contrast as a function of the ring radius $r$ for different integration times $nt_c$. (e through g) Differential photoluminescence counts, contrast, and SNR as a function of the ring width $w$ around the ring radius corresponding to the fluorescence minimum; trace colors follow the notation in (c).

Fig. S4 shows a similar analysis but as a function of the integration time; in all cases we find optimal SNR for $w = 5$ μm (Fig. S4d). Analogously, Fig. S5 displays the impact of the red laser readout power.

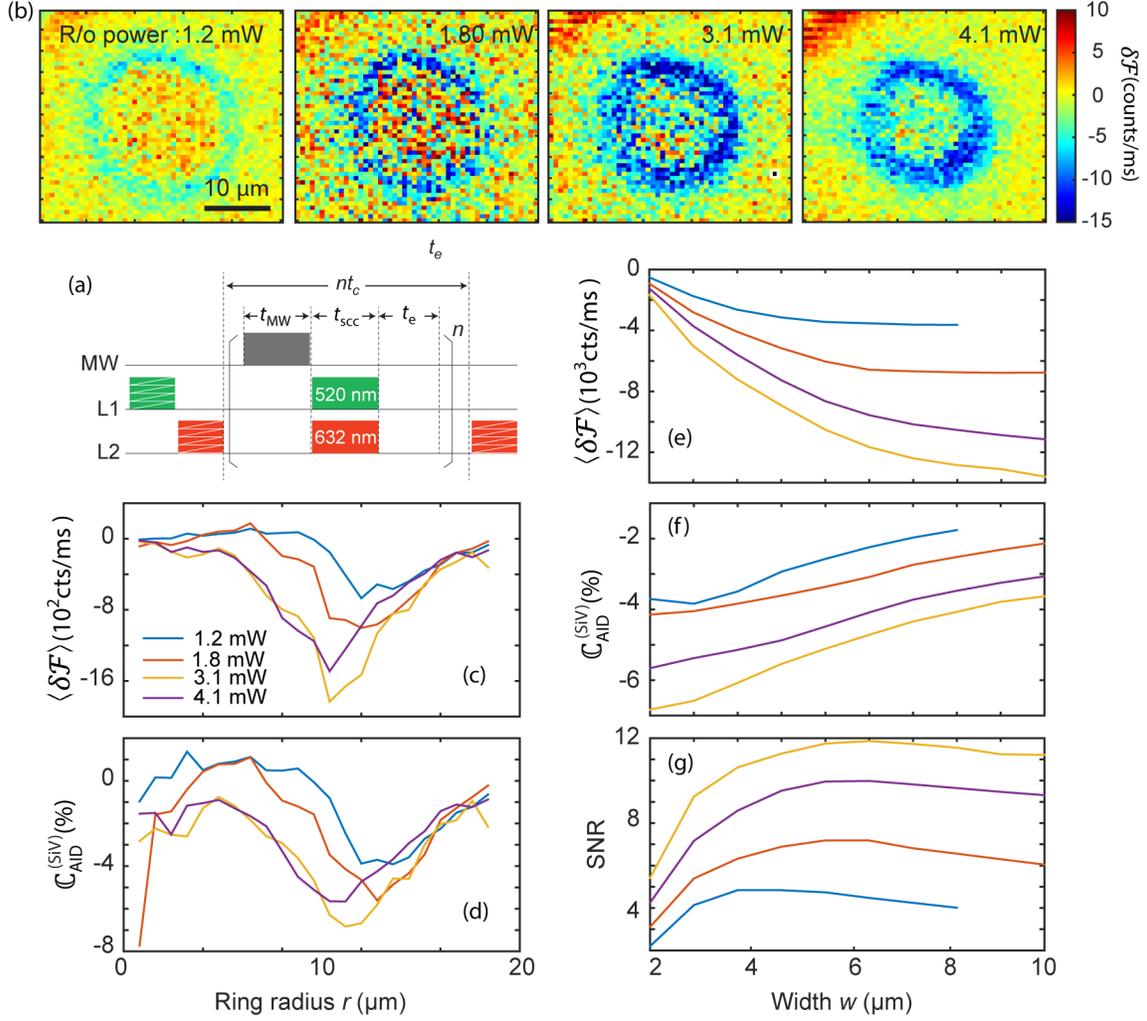

**FIG. S5. Dependence of SiV-AID signal on readout power.** (a) Experimental protocol. In this case, $nt_c = 1$ s, and the readout time per pixel is fixed to 1 ms; all other conditions as in Fig. S3. (b) Differential SiV⁻ photoluminescence images after application of the protocol in (A) for different red laser readout powers (upper right corner in each image). (c through g) Signal counts, contrast, and SNR calculated for varying readout 632 nm laser power as a function of the distance $r$ to the illumination point (for fixed width $w = 1$ μm), or as a function of the ring width $w$ (around a fixed radius $R = 13$ μm).

We find the optimum SNR corresponds to widths slightly above 5 μm, though the change is relatively minor. Finally, Fig. S6 shows the result for the case when neighboring NVs serve as the ancilla traps (Fig. 3 in the main text). Here, we focus on the time buildup of the charge-converted NV disc as a function of the integration time, including the corresponding signal counts and SNR.

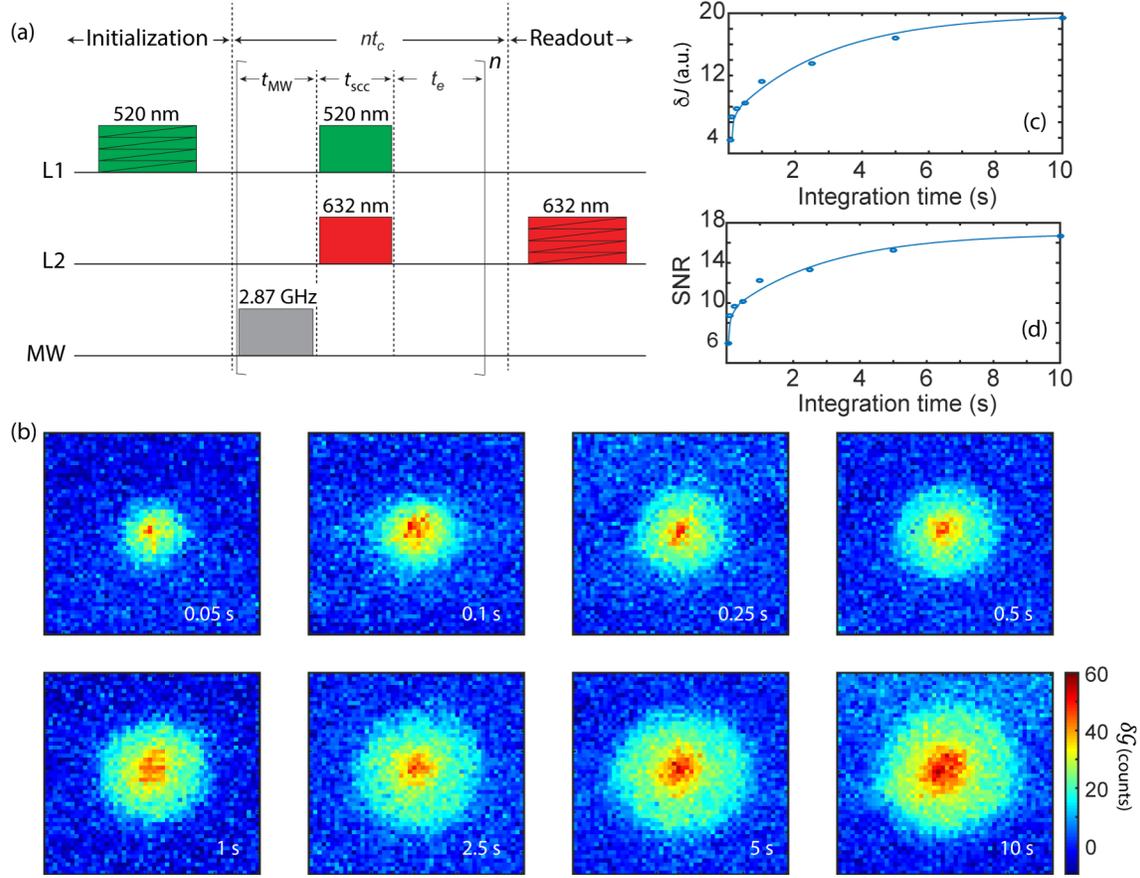

**FIG. S6. NV-AID time buildup.** (a) Experimental protocol. Throughout these experiments, $t_e = t_e = 100$ ns and $t_e = 1$ μs. In this case, the ancilla charge initialization (readout) time is $t'_{ia} = 30$ s ($t'_{ra} = 5$ s); the green laser power is 3 mW and the red laser power is 0.2 mW (21 mW) during readout (SCC). (b) Differential NV⁻ photoluminescence images after application of the protocol in (a) for different integration times $nt_c$ (lower right corners). (c, d) Respectively, signal counts and SNR as a function of the integration time $nt_c$. Solid traces are guides to the eye.